\documentclass{INTERSPEECH2023}
\usepackage{inconsolata}
\usepackage{booktabs} 
\usepackage{pifont}
\usepackage{graphicx} 
\usepackage{multirow}
\usepackage{multicol}
\usepackage{arydshln}
\usepackage{color}
\usepackage{amssymb}
\usepackage{cancel}
\usepackage{diagbox}
\usepackage{makecell}
\usepackage{xcolor}
\usepackage[T1]{fontenc}
\interspeechcameraready 

\usepackage{hyperref}
\hypersetup{
    colorlinks=true
    }
    
\newcommand\blfootnote[1]{%
  \begingroup
  \renewcommand\thefootnote{}\footnote{#1}%
  \addtocounter{footnote}{-1}%
  \endgroup
}

\title{EmoBox: Multilingual Multi-corpus Speech Emotion Recognition \\ Toolkit and Benchmark}
\name{Ziyang Ma$^1$$^\ast$, Mingjie Chen$^2$$^\ast$, Hezhao Zhang$^2$, Zhisheng Zheng$^1$, \\ Wenxi Chen$^1$, Xiquan Li$^1$, Jiaxin Ye$^3$, Xie Chen$^1$$^\dag$, Thomas Hain$^2$$^\dag$}
\address{
  $^1$MoE Key Lab of Artificial Intelligence, X-LANCE Lab, Shanghai Jiao Tong University, China\\
  $^2$Department of Computer Science, University of Sheffield, United Kingdom \\
  $^3$Institute of Science and Technology for Brain-Inspired Intelligence, Fudan University, China}
\email{\{zym.22, chenxie95\}@sjtu.edu.cn, \{mingjie.chen, t.hain\}@sheffield.ac.uk}

\begin{document}
\maketitle
 
\begin{abstract}
\blfootnote{Co-first author$^\ast$. Corresponding author$^\dag$.}
Speech emotion recognition (SER) is an important part of human-computer interaction, receiving extensive attention from both industry and academia. 
However, the current research field of SER has long suffered from the following problems:
1) There are few reasonable and universal splits of the datasets, making comparing different models and methods difficult. 
2) No commonly used benchmark covers numerous corpus and languages for researchers to refer to, making reproduction a burden. 
In this paper, we propose EmoBox\footnote{\url{https://github.com/emo-box/EmoBox}}, an out-of-the-box multilingual multi-corpus speech emotion recognition toolkit, along with a benchmark for both intra-corpus and cross-corpus settings. 
For intra-corpus settings, we carefully designed the data partitioning for different datasets. 
For cross-corpus settings, we employ a foundation SER model, emotion2vec, to mitigate annotation errors and obtain a test set that is fully balanced in speakers and emotions distributions. 
Based on EmoBox, we present the intra-corpus SER results of 10 pre-trained speech models on 32 emotion datasets with 14 languages, and the cross-corpus SER results on 4 datasets with the fully balanced test sets. 
To the best of our knowledge, this is the largest SER benchmark, across language scopes and quantity scales. 
We hope that our toolkit and benchmark can facilitate the research of SER in the community. 

\end{abstract}

\vspace{-0.4cm}
\section{Introduction}
\vspace{-0.1cm}
In the realm of human-computer interaction (HCI), the ability of machines to understand and respond to human emotions through speech has emerged as a pivotal area of research, known as Speech Emotion Recognition (SER). The significance of SER extends across a wide array of applications, from enhancing user experience in virtual assistants~\cite{lin2024advancing} to facilitating emotional well-being in healthcare services~\cite{wu2022climate}. Despite its growing importance, the field of SER faces persistent challenges that hinder progress and innovation. Among these challenges are the scarcity of universally accepted dataset splits~\cite{antoniou2023designing} and the absence of a comprehensive benchmark encompassing a diverse range of corpora and languages~\cite{scheidwasser2022serab}. These limitations complicate the comparison of models and methods, as well as the replication of research findings, thus impeding the advancement of SER technology. 

Recognizing existing critical gaps, this paper introduces EmoBox, a groundbreaking multilingual multi-corpus speech emotion recognition toolkit designed to streamline research in this field. EmoBox is accompanied by a meticulously curated benchmark tailored for both intra-corpus and cross-corpus evaluation settings. For intra-corpus evaluations, we have devised a systematic approach to data partitioning across various datasets, ensuring that researchers can conduct rigorous and comparable analyses of different SER models. In the cross-corpus context, we leverage a foundational SER model, emotion2vec~\cite{ma2023emotion2vec}, to address annotation discrepancies and create a test set that achieves a balance in speaker and emotion distribution, a feat previously unattained in SER research.

\begin{table}[t]
  \centering
  \caption{The datasets involved in EmoBox at a glance.}
  \label{tab:datasets}
  \resizebox{\linewidth}{!}{
  \begin{tabular}{lllllll}
    \toprule
    \textbf{Dataset}  & \textbf{Source} & \textbf{Lang} & \textbf{Emo} & \textbf{Spk} & \textbf{\#Utts} & \textbf{\#Hrs} \\
    \midrule
    AESDD~\cite{vryzas2018speech} & Act & Greek & 5 & 5 & 604 & 0.7 \\
    ASED~\cite{retta2023new} & Act & Amharic & 5 & 65 & 2474 & 2.1 \\
    ASVP-ESD~\cite{landry2020asvp} & Media & Mix & 12 & 131 & 13964 & 18.0 \\
    CaFE~\cite{gournay2018canadian} & Act & French & 7 & 12 & 936  & 1.2\\
    CASIA~\cite{jianhua2008casia} & Act & Mandarin & 6 & 4 & 1200 & 0.6 \\
    CREMA-D~\cite{cao2014crema} & Act & English & 6 & 91 & 7442 & 5.3 \\
    EMNS~\cite{noriy2023emns} & Act & English & 8 & 1 & 1181 & 1.9 \\
    EmoDB~\cite{burkhardt2005database} & Act & German & 7 & 10 & 535 & 0.4\\
    EmoV-DB~\cite{adigwe2018emotional} & Act & English & 5 & 4 & 6887 & 9.5 \\
    EMOVO~\cite{costantini2014emovo} & Act & Italian & 7 & 6 & 588 & 0.5\\
    Emozionalmente~\cite{catania2023speech} & Act & Italian & 7 & 431 & 6902 & 6.3 \\
    eNTERFACE~\cite{martin2006enterface} & Act & English & 6 & 44 & 1263 & 1.1 \\
    ESD~\cite{zhou2021seen} & Act & Mix & 5 & 20 & 35000 & 29.1 \\
    IEMOCAP~\cite{busso2008iemocap} & Act & English & 5 & 10 & 5531 & 7.0\\
    JL-Corpus~\cite{james2018open} & Act & English & 5 & 4 & 2400 & 1.4 \\
    M3ED~\cite{zhao2022m3ed} & TV & Mandarin & 7 & 626 & 24437 & 9.8\\
    MEAD~\cite{wang2020mead} & Act & English & 8 & 48 & 31729 & 37.3 \\
    MELD~\cite{poria2019meld} & TV & English & 7 & 304 & 13706 & 12.1 \\
    MER2023~\cite{lian2023mer} & TV & Mandarin & 6 & / & 5030 & 5.9\\
    MESD~\cite{duville2021mexican} & Act & Spanish & 6 & 11 & 862 & 0.2 \\
    MSP-Podcast~\cite{martinez2020msp} & Podcast & English & 8 & 1273 & 73042 & 113.6 \\
    Oreau~\cite{kerkeni2020french} & Act & French & 7 & 32 & 434 & 0.3 \\
    PAVOQUE~\cite{steiner2013pavoque} & Act & German & 5 & 1 & 7334 & 12.2 \\
    Polish~\cite{miesikowska2020emotions} & Act & Polish & 3 & 5 & 450 & 0.1 \\
    RAVDESS~\cite{livingstone2018ryerson} & Act & English & 8 & 24 & 1440 & 1.5\\
    RESD~\cite{aniemore2022resd} & Act & Russian & 7 & 200 & 1396 & 2.3\\
    SAVEE~\cite{jackson2014savee} & Act & English & 7 & 4 & 480 & 0.5\\
    ShEMO~\cite{mohamad2019shemo} & Act & Persian & 6 & 87 & 2838 & 3.3 \\
    SUBESCO~\cite{sultana2021subesco} & Act & Bangla & 7 & 20 & 7000 & 7.8 \\
    TESS~\cite{dupuis2010tess} & Act & English & 7 & 2 & 2800 & 1.6 \\
    TurEV-DB~\cite{canpolat2020turkish} & Act & Turkish & 4 & 6 & 1735 & 0.5 \\
    URDU~\cite{latif2018cross} & Talk show & Urdu & 4 & 29 & 400 & 0.3 \\
    % CMU-MOSEI~\cite{zadeh2018multimodal} & YouTube & 7 & 1000 & English & 44977 & 91.9\\
    % CMU-MOSI~\cite{zadeh2016mosi} & YouTube & 7 & 89 & English & 2199 & 2.6\\
    % RAVDESS-Song~\cite{livingstone2018ryerson} & Act & 8 & 23 & English & 1012 & 1.3 &\\
    \midrule
    Total & -- & -- & -- & 3510 & 262020 & 294.4 \\
    \bottomrule
  \end{tabular}
  }
  \vspace{-0.7cm}
\end{table}

Our contributions are manifold. Not only do we offer the SER community a powerful toolkit to easily conduct experiments on different datasets, but we also establish a new benchmark for the field. 
We detail our data partitioning, which we believe reduces the burden on researchers for future research. 
We present comprehensive intra-corpus SER results derived from the application of 10 pre-trained speech models across 32 emotion datasets in 14 languages. To our knowledge, this represents the most extensive SER benchmark to date, spanning the broadest scope of languages and the largest scale of data. 
Besides, we showcase the cross-corpus SER performance on 4 datasets, utilizing test sets that are fully balanced in terms of speakers and emotions. Through EmoBox, we aim to provide the SER community with a robust toolkit and benchmark that will catalyze further research, enhance model comparability, and ultimately, foster innovation in the field of speech emotion recognition.

\section{Data Preparation and Partitioning}
\vspace{-0.1cm}
\subsection{Datasets}
\label{sec:datasets}
\vspace{-0.1cm}
The comprehensive overview of the datasets utilized in this study is delineated in Table~\ref{tab:datasets}. There are 32 emotional datasets spanning 14 distinct languages, comprising 12 in English, 3 in Mandarin, and 2 in French, German, and Italian. Additionally, there is 1 dataset each in Amharic, Bangla, Greek, Persian, Polish, Russian, Spanish, Turkish, and Urdu, as well as two datasets featuring a mixture of languages. 

For analytical purposes, each dataset is systematically classified according to several criteria: \textbf{\textit{Source}} denotes the origin of the samples; \textbf{\textit{Lang}} indicates the language of the dataset; \textbf{\textit{Emo}} represents the number of emotional categories encompassed; \textbf{\textit{Spk}} specifies the number of speakers; \textbf{\textit{\#Utts}} details the total number of utterances; and \textbf{\textit{\#Hrs}} quantifies the aggregate hours of the samples.

The speech data extracted from these datasets undergoes a uniform processing protocol, being converted into a monophonic format with a sampling rate of $16,000$ Hz. Each piece of speech data is uniquely annotated with an emotion label, ensuring a precise correlation between the utterance and its emotional categorization. 

\vspace{-0.2cm}
\subsection{Intra-corpus SER}
\vspace{-0.1cm}
Ensuring proper data partitioning is pivotal for leveraging a corpus efficiently, particularly when dealing with corpora of limited size. Through meticulous observation and analysis of the distribution of speakers and emotions across the $32$ datasets, as detailed in Section~\ref{sec:datasets}, we establish a set of criteria for data partitioning as follows:

\begin{enumerate}
    \item Each dataset is divided into training and testing sets, with the division possibly encompassing single or multiple folds depending on the data distribution. 
    \item In cases where datasets come with officially predefined splits, these original partitions are adhered to. For instance, the IEMOCAP dataset is organized into $5$ folds, featuring $2$ distinct speakers per fold, whereas the MELD dataset is split into train, dev, and test splits.
    \item For datasets characterized by a speaker count of fewer than $4$, such as the PAVOQUE dataset which includes only a single speaker; or those with $4$ or more speakers but exhibit an imbalanced distribution of emotions among speakers, such as the M3ED dataset, a speaker-dependent approach is employed. Here, $25\%$ of data for each emotion is earmarked for testing, with the remainder allocated for training.
    \item For datasets whose speaker number is greater than or equal to $4$ with a balanced emotion distribution among speakers, the leave-one-out $n$-fold cross-validation manner is adopted. More specifically, if the number of speakers $\in \{4,5,6\}$, $n$ is set the same as the number of speakers; If the number of speakers exceeds $6$, $n$ is taken to be $4$ and multiple speakers are merged within each fold. 
\end{enumerate}

The data partitioning of the aforementioned criteria is conducted in EmoBox and model performance on the benchmark is thoroughly examined in Section~\ref{sec:intra-corpus-exp}. 
This structured approach to data partitioning underscores our commitment to fostering robust and replicable research methodologies within the field.

\vspace{-0.2cm}
\subsection{Cross-corpus SER}
\vspace{-0.1cm}

\begin{table}[htbp]
  \centering
  \caption{Detailed meta information of different datasets for cross-corpus settings.}
  \label{tab:balanced_meta}
  \resizebox{\linewidth}{!}{
  \begin{tabular}{l|ccc}
  \hline
    & \textbf{Source}   & \textbf{Accent}  & \textbf{Recording} \\
  \hline
  \textbf{IEMOCAP} & Elicitation & English & Crosstalk \\
  \textbf{MELD} & Spontaneousness & English & Noisy \\
  \textbf{RAVDESS}& Act & North American & Clean \\
  \textbf{SAVEE} & Act & British & Clean \\
  \hline
  \end{tabular}
  }
  \vspace{-0.6cm}
\end{table}

In real scenarios, the ability of a SER model to generalize to unseen speakers and unknown recording conditions is essential. 
To evaluate this capability, cross-corpus zero-shot testing emerges as a profitable strategy to assess the robustness of SER models against variability in speakers and recording contexts. 
To implement this approach, we meticulously select  $4$ datasets: IEMOCAP, MELD, RAVDESS, and SAVEE. 
As shown in Table~\ref{tab:balanced_meta}, these datasets cover a diverse range of sources, accents, and recording environments. 
Such diversity is crucial for the cross-corpus setting as it ensures the testing encompasses a broad spectrum of real-world scenarios, thereby providing a comprehensive assessment of models' adaptability and robustness. 

\begin{table}[htbp]
  \centering
  \caption{Balanced test sets statistical information. Each corpus contains 240 pieces of test data, with 4 identical emotions including angry, happy, neutral, and sad.}
  \label{tab:balanced_num}
  \resizebox{\linewidth}{!}{
  \begin{tabular}{l|cccc}
  \hline
    & \textbf{\#Emotion} & \textbf{\#Speaker}   & \textbf{\#Number} & \textbf{\#Total}  \\
  \hline
  \textbf{IEMOCAP} &4 & 10 & 6 &240\\
  \textbf{MELD} &4& 6 & 10 &240\\
  \textbf{RAVDESS} &4& 20 & 3 &240\\
  \textbf{SAVEE} &4& 4 & 15 &240\\
  \hline
  \end{tabular}
  }
  \vspace{-0.5cm}
\end{table}

\begin{table*}[htbp]

  \centering
  \caption{Intra-corpus SER results of 10 pre-trained speech models on 32 emotion datasets spanning 14 distinct languages with EmoBox data partitioning. Unweighted Average Accuracy (UA(\%)), Weighted Average Accuracy (WA(\%)), and Macro F1 Score (F1(\%)) are reported, with \textcolor{red}{Top1}, \textcolor{red!60}{Top2}, and \textcolor{red!20}{Top3} scores highlighted. }
  \resizebox{\linewidth}{!}{
  \begin{tabular}{l|ccc|ccc|ccc|ccc}
  \hline
   \multirow{2}{*}{\textbf{Model}} &  \textbf{UA(\%) $\uparrow$}   & \textbf{WA(\%) $\uparrow$}  & \textbf{F1(\%) $\uparrow$}  & \textbf{UA(\%) $\uparrow$}  & \textbf{WA(\%) $\uparrow$}  & \textbf{F1(\%) $\uparrow$}  & \textbf{UA(\%) $\uparrow$} & \textbf{WA(\%) $\uparrow$}  & \textbf{F1(\%) $\uparrow$} & \textbf{UA(\%) $\uparrow$} & \textbf{WA(\%) $\uparrow$}  & \textbf{F1(\%) $\uparrow$} \\
  \cline{2-13}
   & \multicolumn{3}{c|}{\textbf{AESDD (el)}}   & \multicolumn{3}{c|}{\textbf{ASED (am)}} & \multicolumn{3}{c|}{\textbf{ASVP-ESD (mix)}}  & \multicolumn{3}{c}{\textbf{CaFE (fr)}} \\
  \hline
    wav2vec 2.0 base & 67.59 & 67.65 & 67.61 & 88.59 & 88.63 & 88.63 & 48.99 & 59.12 & 49.78 & 42.76 & 42.47 & 41.03  \\
    HuBERT base & \textcolor{red!60}{82.30}& \textcolor{red!60}{82.35} & \textcolor{red!60}{82.36}  & 94.17 & 94.13 & 94.13 & 48.69 & 59.78 & 49.72 & 54.16 & 54.16 & 53.36\\
    HuBERT large  & {78.85} & {78.90} & {78.88} & 96.19 & 96.19 & 96.17 & \textcolor{red!20}{53.33} & \textcolor{red!20}{63.00} & \textcolor{red!20}{54.14} & \textcolor{red!20}{59.50} & \textcolor{red!20}{58.73} & \textcolor{red!20}{58.22} \\
    WavLM base  & {78.99} & {79.08} &{78.78} & \textcolor{red!20}{94.27} & \textcolor{red!20}{94.31} & \textcolor{red!20}{94.29} & 46.38 & 58.05 & 47.35 & 52.71 & 52.33 & 51.66 	\\
    WavLM large  & \textcolor{red}{84.40} & \textcolor{red}{84.49} & \textcolor{red}{84.19} & \textcolor{red!60}{96.44} & \textcolor{red!60}{96.45} & \textcolor{red!60}{96.42} & \textcolor{red!60}{56.31} & \textcolor{red!60}{65.91} & \textcolor{red!60}{56.83} & \textcolor{red!60}{62.20} & \textcolor{red!60}{61.33} & \textcolor{red!60}{61.14}  \\
    data2vec base  & 49.96 & 50.03 & 49.30 & 86.39 & 86.34 & 86.34 & 37.66 & 50.79 & 38.26 & 42.18 & 42.36 & 41.69 \\
    data2vec large  & 45.63 & 45.69 & 44.65 & 88.31 & 88.31 & 88.30 & 46.95 & 56.36 & 47.33 & 42.24 & 42.85 & 41.11  \\
    data2vec 2.0 base  & 46.55 & 46.68 & 45.33 & 93.99 & 93.99 & 93.98 & 46.00 & 57.57 & 46.62 & 51.83 & 51.67 & 50.52  \\
    data2vec 2.0 large  & 72.26 & 72.34 & 71.82 & 94.30 & 94.28 & 94.27 & 52.18 & 62.35 &	52.64 & 59.04 & 58.02 & 57.51  \\
    Whisper large v3  & \textcolor{red!20}{79.13} & \textcolor{red!20}{79.18} & \textcolor{red!20}{79.13} & \textcolor{red}{96.75} & \textcolor{red}{96.73} & \textcolor{red}{96.74} & \textcolor{red}{61.14} & \textcolor{red}{71.52} & \textcolor{red}{62.08} & \textcolor{red}{69.43} & \textcolor{red}{68.84} & \textcolor{red}{68.06}  \\
  \hline
  \textbf{Model} & \multicolumn{3}{c|}{\textbf{CASIA (zh)}}   & \multicolumn{3}{c|}{\textbf{CREMA-D (en)}} & \multicolumn{3}{c|}{\textbf{EMNS (en)}}  & \multicolumn{3}{c}{\textbf{EmoDB (de)}} \\
  \hline
    wav2vec 2.0 base & 39.56 & 39.56 & 34.86 & 61.95 & 61.90 & 61.75 & 65.14 & 65.27 & 64.80 &82.06& 83.14& 82.21\\
    HuBERT base  & \textcolor{red!20}{47.23} & \textcolor{red!20}{47.23} & \textcolor{red!20}{42.47} & 71.13 & 70.98 & 71.00 & \textcolor{red!60}{75.83} & \textcolor{red!60}{76.25} & \textcolor{red!60}{75.70} & 87.73 & 87.73& 87.82 \\
    HuBERT large & 45.30 & 45.30 & 39.10 & \textcolor{red!20}{73.83} & \textcolor{red!20}{73.64} & \textcolor{red!20}{73.73} & \textcolor{red!20}{73.94} & \textcolor{red!20}{74.28} & \textcolor{red!20}{73.67} &\textcolor{red!20}{89.81}& \textcolor{red!20}{90.26}& \textcolor{red!20}{89.86}   \\
    WavLM base	 & \textcolor{red!20}{47.25} & \textcolor{red!20}{47.25} & \textcolor{red!20}{41.78} & 69.64 & 69.49 & 69.54 & 69.46 & 69.71 & 69.24 & 87.03 & 87.12 & 86.76  \\
    WavLM large  & \textcolor{red!60}{52.12} & \textcolor{red!60}{52.12} & \textcolor{red!60}{46.55} & \textcolor{red!60}{74.50} & \textcolor{red!60}{74.32} & \textcolor{red!60}{74.39} & \textcolor{red}{83.97} & \textcolor{red}{84.12} & \textcolor{red}{83.97} &\textcolor{red}{92.58}& \textcolor{red}{92.67}&\textcolor{red}{92.57} \\
    data2vec base  & 34.72 & 34.72 & 30.88 & 58.03 & 57.78 & 57.73 & 34.33 & 34.58 & 33.12 &58.12& 60.01& 58.32 \\
    data2vec large & 37.65 & 37.65 & 33.50 & 63.80 & 63.51 & 63.48 & 48.52 & 48.96 & 48.39 & 60.96& 61.95& 61.26  \\
    data2vec 2.0 base  & 43.31 & 43.31 & 38.90 & 65.74 & 65.48 & 65.47 & 47.83 & 48.39 & 47.21 & 75.07& 75.86& 75.49\\
    data2vec 2.0 large & 45.57 & 45.57 & 41.46 & 69.55 & 69.27 & 69.25 & 57.80 & 58.60 &	57.15 & 79.36& 80.41& 79.96 \\
    Whisper large v3  & \textcolor{red}{59.58} & \textcolor{red}{59.58} & \textcolor{red}{56.27} & \textcolor{red}{76.75} & \textcolor{red}{76.48} & \textcolor{red}{76.60} & 69.73 & 70.58 & 69.31 &\textcolor{red!60}{91.26}& \textcolor{red!60}{92.43}& \textcolor{red!60}{91.84}\\
  \hline
  \textbf{Model} & \multicolumn{3}{c|}{\textbf{EmoV-DB (en)}}   & \multicolumn{3}{c|}{\textbf{EMOVO (it)}} & \multicolumn{3}{c|}{\textbf{Emozionalmente (it)}}  & \multicolumn{3}{c}{\textbf{eNTERFACE (en)}} \\
  \hline
    wav2vec 2.0 base  & 97.85 & 98.03 & 97.90 &31.07 & 31.07 & 27.24 & 56.69 & 56.69 & 56.64 & 64.19 & 64.12 & 63.81  \\
    HuBERT base  & \textcolor{red!20}{98.72} & \textcolor{red!20}{98.77} & \textcolor{red!20}{98.71} &46.10 & 46.10 & 41.28& 66.30 & 66.30 & 66.26 & 79.14 & 79.11 & 78.97 \\
    HuBERT large & \textcolor{red!60}{99.36} & \textcolor{red!60}{99.40} & \textcolor{red!60}{99.37} & 45.74 & 45.74 & 40.56 & \textcolor{red!20}{69.83} & \textcolor{red!20}{69.83} & \textcolor{red!20}{69.81} & 88.19 & 88.17 & 88.14  \\
    WavLM base	 & 98.38 & 98.49 & 98.39 &42.39 & 42.39 & 37.33 & 63.02 & 63.02 & 63.02 & 88.30 & 88.27 & 88.20 \\
    WavLM large  & \textcolor{red}{99.44} & \textcolor{red}{99.47} & \textcolor{red}{99.45} &\textcolor{red!60}{48.82} & \textcolor{red!60}{48.82} & \textcolor{red!60}{44.16} & \textcolor{red!60}{75.00} & \textcolor{red!60}{75.00} & \textcolor{red!60}{74.97} & \textcolor{red!20}{92.43} & \textcolor{red!20}{92.42} & \textcolor{red!20}{92.40} \\
    data2vec base  & 93.26 & 93.61 & 93.23 &32.47 & 32.47 & 29.22 & 48.97 & 48.97 & 48.77 & 91.61 & 91.62 & 91.64 \\
    data2vec large  & 94.47 & 94.93 & 94.58 &35.66 & 35.66 & 33.59 & 53.92 & 53.92 & 53.66 & 89.46 & 89.46 & 89.65 \\
    data2vec 2.0 base  & 95.81 & 96.09 & 95.80 & 42.96 & 42.96 & 41.01 & 56.22 & 56.22 & 56.00 & 90.81 & 90.80 & 90.83 \\
    data2vec 2.0 large  & 98.17 & 98.32 & 98.19 & \textcolor{red!20}{45.88} & \textcolor{red!20}{45.88} & \textcolor{red!20}{43.63} & 64.10 & 64.10 & 63.93 & \textcolor{red!60}{94.02} & \textcolor{red!60}{94.02} & \textcolor{red!60}{94.01} \\
    Whisper large v3  & \textcolor{red!60}{99.36} & \textcolor{red!60}{99.37} & \textcolor{red!60}{99.34} &\textcolor{red}{57.82}& \textcolor{red}{57.82}& \textcolor{red}{56.06} & \textcolor{red}{76.91} & \textcolor{red}{76.91} & \textcolor{red}{76.90} & \textcolor{red}{97.69} & \textcolor{red}{97.68} & \textcolor{red}{97.68} \\
  \hline
  \textbf{Model} & \multicolumn{3}{c|}{\textbf{ESD (mix)}}   & \multicolumn{3}{c|}{\textbf{IEMOCAP (en)}} & \multicolumn{3}{c|}{\textbf{JL-Corpus (en)}}  & \multicolumn{3}{c}{\textbf{M3ED (zh)}} \\
  \hline
    wav2vec 2.0 base  &69.17 & 69.27 & 68.66& 58.27 & 57.73 & 57.83 & 45.27 & 45.27 & 40.73 &23.13 & 43.20 & 22.91\\
    HuBERT base  &72.41 & 72.41 & 72.11 & 63.87 & 63.10 & 63.45 & 51.11 & 51.11 & 50.56 &23.80 & 42.55 & 24.03\\
    HuBERT large  &75.85 & 75.85 & 75.38 & \textcolor{red!20}{67.42} & \textcolor{red!20}{66.69} & \textcolor{red!20}{67.24} & 56.62 & 56.62 & 52.77 &\textcolor{red!20}{23.25} & \textcolor{red!20}{44.49} & \textcolor{red!20}{23.28} \\
    WavLM base	 &72.90 & 72.90 & 72.55 & 62.92 & 63.94& 63.40 & 53.79 & 53.79 & 52.36 &\textcolor{red!20}{22.76} & \textcolor{red!20}{42.79} & \textcolor{red!20}{22.03} \\
    WavLM large  &\textcolor{red!60}{79.14} & \textcolor{red!60}{79.14} & \textcolor{red!60}{78.87} & \textcolor{red!60}{69.47} & \textcolor{red!60}{69.07} & \textcolor{red!60}{69.29} & \textcolor{red!20}{60.86} & \textcolor{red!20}{60.86} & \textcolor{red!20}{57.34} &\textcolor{red!60}{26.58} & \textcolor{red!60}{44.86} & \textcolor{red!60}{26.98}\\
    data2vec base  &65.05 & 65.05 & 64.55 & 54.19 & 53.20 & 53.76 & 49.29 & 49.29 & 47.86 &19.44 & 37.32 & 19.24 \\
    data2vec large  & 72.01 & 72.01 & 71.77 & 52.56 & 51.11 & 51.71 & 48.87 & 48.87 & 46.92 & 20.20 & 38.73 & 20.26 \\
    data2vec 2.0 base & 73.40& 73.40& 73.10& 54.40 & 53.19 & 53.71 & 54.04 & 54.04 & 52.85 & 22.82 & 41.42 & 22.89  \\
    data2vec 2.0 large & \textcolor{red!20}{76.81}& \textcolor{red!20}{76.81}&\textcolor{red!20}{76.43} & 57.30 & 56.23 & 56.70 & \textcolor{red!60}{65.14} & \textcolor{red!60}{65.14} & \textcolor{red!60}{63.49} &\textcolor{red!20}{23.82} & \textcolor{red!20}{43.02} & \textcolor{red!20}{23.98} \\
    Whisper large v3  & \textcolor{red}{84.62}& \textcolor{red}{84.62} & \textcolor{red}{84.33} & \textcolor{red}{73.54} & \textcolor{red}{72.86} & \textcolor{red}{73.11} & \textcolor{red}{66.71} & \textcolor{red}{66.71} & \textcolor{red}{65.19} & \textcolor{red}{32.84} & \textcolor{red}{49.42} & \textcolor{red}{33.76} \\
  \hline
  \textbf{Model} & \multicolumn{3}{c|}{\textbf{MEAD (en)}}   & \multicolumn{3}{c|}{\textbf{MELD (en)}} & \multicolumn{3}{c|}{\textbf{MER2023 (zh)}}  & \multicolumn{3}{c}{\textbf{MESD (es)}} \\
  \hline
    wav2vec 2.0 base & 72.17 & 73.57 & 72.32 & 20.06 & 45.17 & 20.04 & 40.40 & 46.78 & 40.73 & \textcolor{red!60}{62.93} & \textcolor{red!60}{62.89} & \textcolor{red!60}{62.85}  \\
    HuBERT base  & 74.76 & 75.71 & 74.92 & 23.53 & 45.47 & 24.29 & 42.56 & 49.80 & 42.77 & 47.52 & 47.48 & 46.33  \\
    HuBERT large  & \textcolor{red!60}{76.87} & \textcolor{red!60}{77.84} & \textcolor{red!60}{77.12} & 24.13 & 46.37 & 24.99 & \textcolor{red!20}{43.96} & \textcolor{red!20}{50.49} & \textcolor{red!20}{44.45} & 53.71 & 53.67 & 53.67  \\
    WavLM base  & 71.85 & 72.86 & 72.14 & 23.44 & 44.71 & 24.25 & 41.80 & 48.71 & 41.97 & 43.58 & 43.52 & 42.94 	\\
    WavLM large  &  \textcolor{red}{81.27} &  \textcolor{red}{82.03} &  \textcolor{red}{81.43} & \textcolor{red!60}{28.18} & \textcolor{red!60}{49.31} & \textcolor{red!60}{29.11} & \textcolor{red!60}{48.17} & \textcolor{red!60}{54.77} & \textcolor{red!60}{49.36} & \textcolor{red!20}{62.54} & \textcolor{red!20}{62.49} & \textcolor{red!20}{62.33} \\
    data2vec base  & 65.36 & 66.05 & 65.53 & 23.82 & 45.57 & 24.37 & 37.94 & 43.06 & 38.15 & 34.37 & 34.35 & 33.24  \\
    data2vec large  & 67.57 & 68.44 & 67.88 & 23.35 & 45.74 & 24.10 & 35.14 & 40.27 & 34.28 & 36.67 & 36.61 & 35.81  \\
    data2vec 2.0 base  & 69.66 & 70.64 & 69.90 & 24.79 & 46.65 & 25.28 & 42.59 & 46.64 & 43.22 & 44.86 & 44.85 & 43.6 \\
    data2vec 2.0 large  & 74.13 & 75.24 & 74.43 & \textcolor{red!20}{26.33} & \textcolor{red!20}{47.72} & \textcolor{red!20}{27.35} & 42.05 & 46.81 & 42.08 & 48.46 & 48.45 & 46.8 \\
    Whisper large v3  & \textcolor{red!20}{76.35} & \textcolor{red!20}{77.34} & \textcolor{red!20}{76.55} & \textcolor{red}{31.54} & \textcolor{red}{51.89} & \textcolor{red}{32.95} & \textcolor{red}{61.22} & \textcolor{red}{65.23} & \textcolor{red}{62.29} & \textcolor{red}{69.78} & \textcolor{red}{69.67} & \textcolor{red}{69.64} \\
  \hline
   \textbf{Model} & \multicolumn{3}{c|}{\textbf{MSP-Podcast (en)}}   & \multicolumn{3}{c|}{\textbf{Oreau (fr)}} & \multicolumn{3}{c|}{\textbf{PAVOQUE (de)}}  & \multicolumn{3}{c}{\textbf{Polish (pl)}} \\
  \hline
    wav2vec 2.0 base  & 15.5& 37.48& 13.8& 40.14 & 41.06 & 39.23 & 84.95 & 91.23 & 86.09 & 67.35 & 67.35 & 66.76 \\
    HuBERT base  & 16.97& 38.7& 15.89 & 51.98 & 52.80 & 51.89 & 86.12 & 92.19 & 87.06 & 69.40 & 69.40 & 69.08 \\
    HuBERT large  & 18.07 & 40.02 & 17.2 & 63.69 & 64.35 & 63.66 & \textcolor{red!20}{87.04} & \textcolor{red!20}{92.76} & \textcolor{red!20}{87.94} & 70.20 & 70.20 & 70.74 \\
    WavLM base	 & 17.11 & 37.38 & 16.53 & 57.78 & 58.06 & 57.45 & 83.73 & 90.84 & 85.60 & 69.31 & 69.31 & 69.46 \\
    WavLM large  & \textcolor{red!60}{18.6} & \textcolor{red!60}{40.47} & \textcolor{red!60}{17.97} & \textcolor{red!60}{65.54} & \textcolor{red!60}{66.29} & \textcolor{red!60}{65.67} & \textcolor{red}{87.73} & \textcolor{red}{93.40} & \textcolor{red}{88.43} &  \textcolor{red!60}{79.29} &  \textcolor{red!60}{79.29} &  \textcolor{red!60}{79.02} \\
    data2vec base  & 16.19 & 37.15 & 15.49 & 54.84 & 55.67 & 54.76 & 74.94 & 85.11 & 76.63 & 71.31 & 71.31 & 70.65 \\
    data2vec large  & 17.24 & 37.45 & 16.86 & 48.29 & 49.12 & 48.21 & 73.26 & 84.89 & 75.71 & 68.05 & 68.05 & 67.07  \\
    data2vec 2.0 base  & 16.79 & 39.97 & 15.33 & 59.40 & 59.73 & 58.13 & 78.3 & 87.82 & 80.92 & \textcolor{red!20}{75.75} & \textcolor{red!20}{75.75} & \textcolor{red!20}{75.52}  \\
    data2vec 2.0 large  &  \textcolor{red!20}{18.35} &  \textcolor{red!20}{41.33} &  \textcolor{red!20}{17.07} & \textcolor{red!20}{64.64} & \textcolor{red!20}{64.88} & \textcolor{red!20}{64.50} & 85.38 & 92.07 & 86.75 & 74.00 & 74.00 & 74.05 \\
    Whisper large v3  & \textcolor{red}{22.24} & \textcolor{red}{44.1} & \textcolor{red}{22.12} & \textcolor{red}{84.48} & \textcolor{red}{84.79} & \textcolor{red}{85.01} & \textcolor{red!60}{87.72} & \textcolor{red!60}{93.17} & \textcolor{red!60}{88.41} & \textcolor{red}{83.27} & \textcolor{red}{83.27} & \textcolor{red}{82.77} \\
  \hline
   \textbf{Model} & \multicolumn{3}{c|}{\textbf{RAVDESS (en)}}   & \multicolumn{3}{c|}{\textbf{RESD (ru)}} & \multicolumn{3}{c|}{\textbf{SAVEE (en)}}  & \multicolumn{3}{c}{\textbf{ShEMO (fa)}} \\
  \hline
    wav2vec 2.0 base  & 54.33 & 55.40 & 53.99 &\textcolor{red!20}{52.82}& \textcolor{red!20}{53.39}& \textcolor{red!20}{52.90} &44.89& 49.83& 42.07 & 56.34 & 78.96 & 57.34 \\
    HuBERT base  & 65.43 & 66.21 & 65.31 & 51.56& 52.34& 51.65&58.90 &59.85& 64.05& 58.31 & 81.17 & 63.15 \\
    HuBERT large  & 70.00 & 70.29 & 69.54 &50.82& 51.51& 50.74&71.91& 75.05& 71.83 & \textcolor{red!20}{64.29} & \textcolor{red!20}{83.35} & \textcolor{red!20}{66.26} \\
    WavLM base	 & 61.56 & 62.10 & 61.18 & 44.81& 45.09& 44.97 &63.57& 67.05& 62.83& 60.73 & 78.76 & 62.06 \\
    WavLM large  & \textcolor{red!60}{72.00} & \textcolor{red!60}{72.22} & \textcolor{red!60}{71.42}& \textcolor{red}{55.87}& \textcolor{red}{56.47}& \textcolor{red}{55.82} &66.74& 70.80& 66.37 & \textcolor{red!60}{71.72} & \textcolor{red!60}{87.13} & \textcolor{red!60}{73.55} \\
    data2vec base  & 51.92 & 52.22 & 51.10 & 37.09& 37.90& 36.86 &\textcolor{red}{75.65}& \textcolor{red}{78.25}& \textcolor{red}{78.38} & 47.61 & 70.07 & 49.49 \\
    data2vec large  & 59.30 & 59.50 & 58.74 & 30.78& 31.57& 30.81 &68.01& 71.45& 71.08& 56.42 & 74.09 & 60.98 \\
    data2vec 2.0 base  & 64.66 & 64.84 & 64.18 & 43.54& 44.03& 43.00 &\textcolor{red!20}{72.65}& \textcolor{red!20}{75.50}& \textcolor{red!20}{75.59}& 60.59 & 79.03 & 64.03 \\
    data2vec 2.0 large  & \textcolor{red!20}{71.15} & \textcolor{red!20}{71.63} & \textcolor{red!20}{70.94} & 44.08& 44.64& 44.25 &\textcolor{red}{75.75}& \textcolor{red}{78.59}& \textcolor{red}{78.24} & \textcolor{red!20}{64.09} & \textcolor{red!20}{82.68} & \textcolor{red!20}{68.47} \\
    Whisper large v3  & \textcolor{red}{75.32} & \textcolor{red}{75.87} & \textcolor{red}{75.19} & \textcolor{red!60}{54.98}& \textcolor{red!60}{55.54}& \textcolor{red!60}{54.99} & \textcolor{red!60}{74.07}& \textcolor{red!60}{77.24}& \textcolor{red!60}{75.31}& \textcolor{red}{80.23} & \textcolor{red}{89.55} & \textcolor{red}{82.94} \\
  \hline
   \textbf{Model} & \multicolumn{3}{c|}{\textbf{SUBESCO (bn)}}   & \multicolumn{3}{c|}{\textbf{TESS (en)}} & \multicolumn{3}{c|}{\textbf{TurEV-DB (tr)}}  & \multicolumn{3}{c}{\textbf{URDU (ur)}} \\
  \hline
    wav2vec 2.0 base & 51.25 & 51.25 & 50.91 & 97.92 & 97.92 & 97.92 & 67.19 & 68.01 & 67.19 & \textcolor{red!60}{87.50} & \textcolor{red!60}{87.50} & \textcolor{red!60}{87.57} \\
    HuBERT base &  57.89 & 57.89 & 57.64 & 99.62 & 99.62 & 99.62 & \textcolor{red!20}{72.81} & \textcolor{red!20}{73.26} & \textcolor{red!20}{72.89} & \textcolor{red}{88.41} & \textcolor{red}{88.41} & \textcolor{red}{88.40}  \\
    HuBERT large & 64.53 & 64.53 & 64.33 & \textcolor{red!60}{99.86} & \textcolor{red!60}{99.86} & \textcolor{red!60}{99.86} & 70.93 & 72.08 & 70.96 & 81.75 & 81.75 & 81.66 \\
    WavLM base & 57.06 & 57.06 & 56.78 & 99.10 & 99.10 & 99.10 & 69.97 & 70.51 & 70.18 & 82.82 & 82.82 & 82.85\\
    WavLM large & \textcolor{red!20}{65.33} & \textcolor{red!20}{65.33} & \textcolor{red!20}{65.09} & \textcolor{red!20}{99.78} & \textcolor{red!20}{99.78} & \textcolor{red!20}{99.78} & \textcolor{red!60}{79.5} & \textcolor{red!60}{80.09} & \textcolor{red!60}{79.51} & \textcolor{red!20}{86.61} & \textcolor{red!20}{86.61} & \textcolor{red!20}{86.64}\\ 
    data2vec base & 46.22 & 46.22 & 45.80 & 94.67 & 94.67 & 94.65 & 51.62 & 52.58 & 51.52 & 65.42 & 65.42 & 65.35 \\
    data2vec large & 53.06 & 53.06 & 52.62 & 96.89 & 96.89 & 96.89 & 54.69 & 55.11 & 54.56 & 64.81 & 64.81 & 64.93 \\
    data2vec 2.0 base & 57.97 & 57.97 & 57.69 & 98.55 & 98.55 & 98.55 & 63.44 & 63.69 & 63.37 & 69.97 & 69.97 & 69.91 \\
    data2vec 2.0 large & \textcolor{red!60}{66.45} & \textcolor{red!60}{66.45} & \textcolor{red!60}{66.25} & 99.54 & 99.54 & 99.54 & 63.77 & 64.13 & 63.00 & 78.10 & 78.10 & 78.10 \\
    Whisper large v3 & \textcolor{red}{73.05} & \textcolor{red}{73.05} & \textcolor{red}{72.94} & \textcolor{red}{99.96} & \textcolor{red}{99.96} & \textcolor{red}{99.96} & \textcolor{red}{81.32} & \textcolor{red}{81.58} & \textcolor{red}{81.31} & 82.52 & 82.52 & 82.41 \\
  \hline
    
  \end{tabular}
  }
  \label{tab:intra-corpus}
\end{table*}

To address potential errors in annotation, we leverage the fine-tuned version\footnote{\url{https://github.com/ddlBoJack/emotion2vec}} of the emotion2vec~\cite{ma2023emotion2vec} model, a foundation speech emotion recognition model with iterative fine-tuning on over $10,000$ hours of speech data. 
Our methodology involves an initial application of emotion2vec to assign pseudo-labels to our datasets. Subsequently, we refine our dataset by retaining only those instances where there is congruence between the original annotations and those generated by emotion2vec. This step ensures the reduction of annotation discrepancies and enhances the reliability for further analysis.
To establish a fully balanced test set across each dataset, we select $240$ speech-emotion pairs for each dataset. Shared emotions among these datasets contain angry, happy, neutral, and sad, resulting in $60$ pieces for each emotion. 
The composition of the test set, including the number of speakers and the allocation of speech-emotion pairs per speaker for each emotion, is detailed in Figure~\ref{tab:balanced_num}. 
For the IEMOCAP and SAVEE datasets, all speakers are encompassed, including $5$ male and $5$ female speakers in IEMOCAP, and $4$ male speakers in SAVEE. In the case of the MELD dataset, we focus on the $6$ protagonists. For the RAVDESS dataset, we include a selection of 20 speakers, equally divided between $10$ male and $10$ female speakers.
This meticulous composition of the test set is instrumental in our analysis of the models' generalization capabilities across diverse corpora, helping our assessment of the robustness and adaptability of models under cross-corpus settings. 

\section{Benchmark}

\subsection{Experiments Setup}
\vspace{-0.1cm}
$10$ pre-trained models are employed to establish the benchmark of EmoBox, including self-supervised wav2vec 2.0~\cite{baevski2020wav2vec} base, HuBERT~\cite{hsu2021hubert} base/large, WavLM~\cite{chen2022wavlm} base/large, data2vec~\cite{baevski2022data2vec} base/large, data2vec 2.0~\cite{baevski2023efficient} base/large, and an additional supervised ASR encoder of Whisper~\cite{radford2023robust} large v3. 
Speaking of the parameters, the base SSL model is about 95M, the large SSL model is about 315M, and the Whisper encoder is the largest, for about 635M. 
We extract features of the pre-trained models from the last Transformer layer~\footnote{We found that the last layer features from wav2vec 2.0 large fail to conduct SER task on the most corpus, so we did not report them.} and conduct layer normalization uniformly to speed up convergence.
The downstream networks are a simple linear hidden layer and a classification head with a ReLU activation function and a pooling layer sandwiched between them. 
For each dataset and each model, we sweep the learning rate $\in \{1e-3, 1e-4\}$ and hidden layer dimensions $\in \{128, 256\}$. The optimal settings, determined by the best performance outcomes, are then selected for presentation. 
In each training, $20\%$ of data from the training set is selected as the validation set. 
All experiments are trained for $100$ epochs, with the first $10$ epochs to warm up to the maximum learning rate. 
Three key performance metrics are reported: Unweighted Average Accuracy (UA), Weighted Average Accuracy (WA), and Macro F1 Score in the evaluation of the benchmark of EmoBox. 

\vspace{-0.3cm}
\subsection{Intra-corpus SER Results}
\label{sec:intra-corpus-exp}
\vspace{-0.1cm}

Table~\ref{tab:intra-corpus} presents intra-corpus SER results of selected $10$ pre-trained speech models on $32$ emotion datasets spanning $14$ distinct languages, with EmoBox data partitioning. 
The experimental results show that the Whisper large v3 encoder performs significantly better than other SSL models, ranking top$1$ on $23/32$ datasets and top$3$ on $30/32$ datasets. 
The possible reason is that whisper large v3 is trained with speech from multiple languages with data and model scaling. 
Except for Whisper large v3 encoder, WavLM large performs best, ranking top$3$ on $31/32$ datasets, followed by HuBERT large with $14/32$ and data2vec 2.0 large with $12/32$. 
All three models are pre-trained on English speech data, while WavLM employs more data as well as introduces noise in SSL training to enhance robustness. 
This suggests that SSL features trained with more data can enhance the performance of SER. 
Researchers can explore more about emotion in speech with EmoBox benchmark, such as linguistic correlation and difference. 

\vspace{-0.3cm}
\subsection{Cross-corpus SER Results}
\vspace{-0.1cm}

Table~\ref{tab:cross_corpus_each} presents cross-corpus SER results of selected $10$ pre-trained speech models on the refined EmoBox test sets. 
As seen from the table, Whisper large v3 encoder still performs best on the cross-corpus settings, with top$1$ on $9/12$ train-test pairs, while HuBERT large, WavLM large and data2vec base takes $1$ each, respectively. 

\begin{figure}[htbp]
    \centering
    \includegraphics[width=0.8\linewidth]{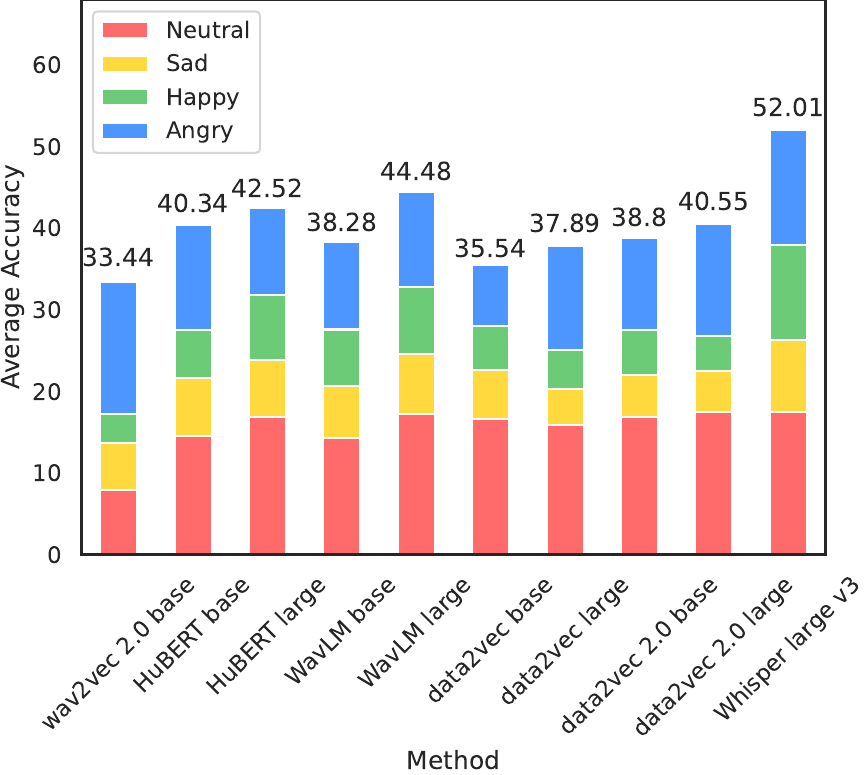}
    \vspace{-0.2cm}
    \caption {Average accuracy for cross-corpus settings.}
    \label{fig:cross-corpus}
    \vspace{-0.4cm}
\end{figure}

Figure~\ref{fig:cross-corpus} illustrates the average accuracy of different models with cross-corpus settings. 
The formula for calculating the average accuracy $\overline{Acc}$ is given as follows:

\begin{equation}
\overline{Acc} = \frac{1}{n^2 - n} \sum_{\substack{i=1 \\ i \neq j}}^{n} \sum_{j=1}^{n} Acc_{i,j},
\end{equation}
where $Acc_{i,j}$ denotes training with dataset $i$ and testing on dataset $j$, and $n=4$ is in our settings. 
As seen from Figure~\ref{fig:cross-corpus}, Except for Whisper large v3 encoder, WavLM performs best among large model sizes and HuBERT base performs best among base model sizes. 

\vspace{-0.2cm}
\section{Conclusion, Limitation and Future Work}
EmoBox offers an easy-to-use toolkit for multilingual multi-corpus SER research with data preparation and partitioning, and the largest benchmark for intra-corpus and cross-corpus evaluations to date. 
We invite the research community to adopt EmoBox toolkit and evaluate EmoBox benchmark for advancing SER methodologies, thereby contributing to the development of more emotionally intelligent human-computer interactions. 
Due to the huge workload, we only evaluate the last layer of features from the pretrained model. 
We will design more comprehensive evaluations in the future and give more instructive generalization conclusions to the SER community. 

\begin{table}[htbp]
  \centering
  \caption{Accuracy (\%) with features from different pre-training models for cross-corpus settings, where the horizontal direction represents the training sets, while the vertical direction represents the test sets. \textbf{I}, \textbf{M} \textbf{R}, \textbf{S} stand for IEMOCAP, MELD, RAVDESS and SAVEE, respectively. \textbf{BOLD} indicates the best results for each train-test pair among 10 pre-trained models.}
  \label{tab:cross_corpus_each}
  \resizebox{\linewidth}{!}{
  \begin{tabular}{l|cccc|cccc}
  \hline
  \multirow{2}{*}{} & \textbf{I} & \textbf{M}   & \textbf{R} & \textbf{S}   & \textbf{I} & \textbf{M}   & \textbf{R} & \textbf{S}\\
  \cline{2-9}
  & \multicolumn{4}{c|}{\textbf{wav2vec 2.0 base}}   & \multicolumn{4}{c}{\textbf{Whisper large v3}} \\
  \hline
  \textbf{I} & \diagbox[dir=SE]{}{} & 29.78 & 18.25 & 28.84 & \diagbox[dir=SE]{}{} & 46.14 & 38.24 & \textbf{46.12} \\
  \textbf{M} & 22.50 & \diagbox[dir=SE]{}{} & 31.39 & 35.24 & \textbf{51.42} & \diagbox[dir=SE]{}{} & \textbf{47.00} & 36.44 \\
  \textbf{R} & 27.15 & 23.20 & \diagbox[dir=SE]{}{} & 33.77 & \textbf{48.12} & \textbf{40.68} & \diagbox[dir=SE]{}{} & \textbf{66.91} \\
  \textbf{S} & 31.34 & 29.19 & 21.36 &\diagbox[dir=SE]{}{} & \textbf{49.30} & \textbf{42.18} & \textbf{49.63}   &\diagbox[dir=SE]{}{}  \\
  \hline
  & \multicolumn{4}{c|}{\textbf{HuBERT base}}   & \multicolumn{4}{c}{\textbf{HuBERT large}} \\
  \hline
  \textbf{I} & \diagbox[dir=SE]{}{} & 37.32 & 22.63 & 35.88 & \diagbox[dir=SE]{}{} & 44.60 & 15.03 & 39.99 \\
  \textbf{M} & 38.31 & \diagbox[dir=SE]{}{} & 31.60 & 32.67 & 44.69 & \diagbox[dir=SE]{}{} & 38.22 & \textbf{43.74} \\
  \textbf{R} & 42.00 & 33.43 & \diagbox[dir=SE]{}{} & 43.47 & 36.18 & 25.02 & \diagbox[dir=SE]{}{} & 56.96 \\
  \textbf{S} & 39.39 & 29.03 & 38.41 & \diagbox[dir=SE]{}{} & 42.81 & 31.54 & 31.92 & \diagbox[dir=SE]{}{}  \\
  \hline
  & \multicolumn{4}{c|}{\textbf{WavLM base}}   & \multicolumn{4}{c}{\textbf{WavLM large}} \\
  \hline
  \textbf{I} & \diagbox[dir=SE]{}{} & 38.25 & 27.80 & 39.40 & \diagbox[dir=SE]{}{} & \textbf{48.59} & 34.16 & 35.53 \\
  \textbf{M} & 46.30 & \diagbox[dir=SE]{}{} & 21.38 & 35.75 & 39.06 & \diagbox[dir=SE]{}{} & 23.06 & 25.74 \\
  \textbf{R} & 30.78 & 29.58 & \diagbox[dir=SE]{}{} & 43.24 & 44.03 & 33.90 & \diagbox[dir=SE]{}{} & 63.35 \\
  \textbf{S} & 34.00 & 27.40 & 27.38 &\diagbox[dir=SE]{}{} & 43.69 & 34.10 & 36.69 &\diagbox[dir=SE]{}{}  \\
  \hline
  & \multicolumn{4}{c|}{\textbf{data2vec base}}   & \multicolumn{4}{c}{\textbf{data2vec large}} \\
  \hline
  \textbf{I} & \diagbox[dir=SE]{}{} & 43.86 & \textbf{40.46} & 32.53 &  \diagbox[dir=SE]{}{} & 44.99 & 39.03 & 36.88 \\
  \textbf{M} & 42.57 & \diagbox[dir=SE]{}{} & 24.16 & 24.33 & 40.62 & \diagbox[dir=SE]{}{} & 29.10 & 31.57 \\
  \textbf{R} & 22.28 & 20.32 & \diagbox[dir=SE]{}{} & 27.02& 26.50 & 26.73 & \diagbox[dir=SE]{}{} & 32.54\\
  \textbf{S} & 29.41 & 31.96 & 34.68 &\diagbox[dir=SE]{}{} & 26.82 & 24.05 & 16.96 &\diagbox[dir=SE]{}{}  \\
  \hline
  & \multicolumn{4}{c|}{\textbf{data2vec 2.0 base}}   & \multicolumn{4}{c}{\textbf{data2vec 2.0 large}} \\
  \hline
  \textbf{I} & \diagbox[dir=SE]{}{} & 44.96 & 30.52 & 31.40 &  \diagbox[dir=SE]{}{} & 47.43 & 17.80 & 29.67 \\
  \textbf{M} & 42.35 & \diagbox[dir=SE]{}{} & 33.32 & 30.42 & 41.75 & \diagbox[dir=SE]{}{} & 31.80 & 29.66 \\
  \textbf{R} & 30.77 & 26.80 & \diagbox[dir=SE]{}{} & 37.31 & 38.79 & 34.21 & \diagbox[dir=SE]{}{} & 35.43 \\
  \textbf{S} & 29.12 & 35.29 & 19.24 &\diagbox[dir=SE]{}{} & 36.39 & 37.79 & 23.58 &\diagbox[dir=SE]{}{}  \\
  \hline
  \end{tabular}
  }
  \vspace{-0.5cm}
\end{table}

% \begin{table}[htbp]
%   \centering
%   \caption{Average scores for cross-corpus settings.}
%   \label{tab:cross_corpus_average}
%   \resizebox{\linewidth}{!}{
%   \begin{tabular}{l|ccccc}
%   \hline
%     & \textbf{Angry} & \textbf{Happy} & \textbf{Neutral} & \textbf{Sad}  & \textbf{Average}  \\
%   \hline
%     wav2vec 2.0 base \\
%     HuBERT base \\
%     HuBERT large \\
%     WavLM base	\\
%     WavLM large \\
%     data2vec base \\
%     data2vec large \\
%     data2vec 2.0 base \\
%     data2vec 2.0 large \\
%     Whisper large v3 \\
%   \hline
%   \end{tabular}
%   }
% \end{table}

\vspace{-0.3cm}
\section{Acknowledgements}
This work was supported by the National Natural Science Foundation of China  (No. 62206171 and No. U23B2018), Shanghai Municipal Science and Technology Major Project under Grant 2021SHZDZX0102 and the International Cooperation Project of PCL.
This work was also supported by Liveperson, Inc. and conducted at the Voicebase/Liveperson Centre of Speech and Language Technology at the University of Sheffield.

\newpage
\bibliographystyle{IEEEtran}
\bibliography{mybib}

\end{document}